\documentclass[fleqn,usenatbib]{mnras}

\usepackage[T1]{fontenc}
\usepackage{ae,aecompl}

\usepackage{graphicx}	
\usepackage{amsmath}	
\usepackage{amssymb}	
\usepackage{comment}    
\usepackage{float}
\usepackage{epstopdf}
\usepackage[colorinlistoftodos]{todonotes}  %

\title[Temperature fluctuations in Seyfert galaxies]{Electron temperature fluctuations in Seyfert galaxies}
\author[R. A. Riffel et al.]{
Rogemar A. Riffel,$^{1}$\thanks{E-mail: rogemar@ufsm.br}
Oli L. Dors,$^{2}$ 
Angela C. Krabbe,$^{2}$
C\'esar Esteban$^{3,4}$
\\
$^{1}$Departamento de F\'isica, Centro de Ci\^encias Naturais e Exatas, Universidade Federal de Santa Maria, 97105-900, Santa Maria, RS, Brazil\\ 
$^{2}$Universidade do Vale do Para\'iba, Av. Shishima Hifumi, 2911, Cep
12244-000, S\~ao Jos\'e dos Campos, SP, Brazil \\
$^{3}$ Instituto de Astrof\'isica de Canarias (IAC), E-38205 La Laguna, Spain\\
$^{4}$ Departamento de Astrof\'isica, Universidad de La Laguna, E-38206 La Laguna, Spain\\
}

\date{Accepted XXX. Received YYY; in original form ZZZ}

\pubyear{2021}

\begin{document}
\label{firstpage}
\pagerange{\pageref{firstpage}--\pageref{lastpage}}
\maketitle

\begin{abstract}
We use Gemini GMOS-IFU observations of three luminous nearby Seyfert galaxies (Mrk\,79, Mrk\,348 and Mrk\,607) to estimate the electron temperature ($T_{\rm e}$) fluctuations in the  inner 0.4--1.1 kpc  region of these galaxies. Based on $T_{\rm e}$
determinations through the [\ion{O}{iii}]$\lambda5007$/$\lambda4363$ emission line ratio of each spaxel,  temperature variations are quantified by  computing the integrated value of the temperature fluctuation parameter ($t^{\rm 2}$) projected in the plane of the sky $t_{\rm A}^{\rm 2}$, for the first time in Active Galactic Nuclei.  We find $t_{\rm A}^{\rm 2}$ values of 0.135, 0.039, and 0.015 for Mrk\,79, Mrk\,348, and Mrk\,607, respectively, which are of the same order or larger than the maximum values reported in star-forming regions and planetary nebulae. Taking into account that $t_{\rm A}^{\rm 2}$ should be considered a lower limit of the total $t^2$ in the nebular volume, the results suggest that the impact of such fluctuations on chemical abundance determinations can be important in some AGNs.
\end{abstract}




\section{Introduction}
\label{introduction}
Metal abundance determinations in the gas phase of line emitting objects, such as star-forming regions
(SFs),
Planetary Nebulae (PNs) and Active Galactic Nuclei (AGNs), play a key role in  the studies of chemical enrichment of the  interstellar medium (ISM) and of the cosmic evolution of the Universe.
Therefore, along decades several efforts have been made in developing methods to derive the chemical abundance and metallicity in these objects.

There is consensus that, for SFs and PNs, reliable metal  abundance determinations
are those based on direct determination of the electron temperature ($T_{\rm e}$), which is obtained through the  measurements of auroral lines sensitive to $T_{\rm e}$, for instance, [\ion{O}{iii}]$\lambda$4363,
[\ion{N}{ii}]$\lambda$5755 and [\ion{S}{iii}]$\lambda$6312. This approach is called the $T_{\rm e}$-method (for a review see \citealt{peimbert17,enrique17}) and it has been used to directly determine the chemical abundance of thousands of local SFs (e.g. \citealt{smith75,torrespeimber89,vanzee98,kennicutt03,hagele08,zurita12,berg20}, among others)
as well as for some objects at high redshifts ($z \: > \: 1$; e.g.\ \citealt{sanders16,sanders20,gburek19}).
There is considerable observational evidence that supports the reliability of $T_{\rm e}$-method, such as, for example, the agreement between oxygen abundances obtained for SFs located in the solar neighborhood and those obtained from observations of the weak interstellar \ion{O}{i}$\lambda$1356 line towards stars (see \citealt{pilyugin03} and references therein), as well as between oxygen abundances obtained for SFs and B-type stars in the Milky Way and other nearby galaxies (see \citealt{toribio17} and references therein).

However, abundance determinations using the $T_{\rm e}$-method are subject to some caveats. It is worth to state from the onset that, \citet{peimbert67} proposed  the presence of 
 electron temperature fluctuations in \ion{H}{ii} regions. In  that work, temperatures based on collisionally excited lines (CELs), i.e.  [\ion{O}{iii}]($\lambda$4949+$\lambda$5007)/$\lambda$4363, from the SFs M\,8, M\,17 and Orion Nebulae were found to be 2000-4000 K higher than those determined from the Balmer continuum and radio observations (see also \citealt{2007ApJ...670..457G}). Thereafter, \citet{peimbert69}  reported that temperature variations over the observed volume of the nebula, if not taken into account, produce underestimated gaseous abundances derived through intensity ratios of CELs using the $T_{\rm e}$-method.

In fact, carbon and oxygen abundances derived from faint 
optical recombination lines (ORLs) (almost independent from $T_{\rm e}$) of SFs are higher by a factor of 0.2-0.4 dex than those derived from the $T_{\rm e}$-method (e.g. \citealt{tsamis03,esteban04,peimbert05,garciarojas06}). This is known as the abundance discrepancy problem and several scenarios have been proposed to explain it (for a review see \citealt{2007ApJ...670..457G}).
If this discrepancy is due to presence of electron temperature fluctuations (quantified by the $t^{2}$ parameter),
as suggested by \citet{peimbert69}, large variations
of $T_{\rm e}$ along  gaseous nebulae could be derived to a level of $t^{2}\sim0.04$ (e.g. \citealt{2007ApJ...670..457G}). However, observations of SFs  over decades have failed to find direct evidence for such high levels of electron temperature fluctuations. 
A discrete estimation of $t^2$ in the plane of the sky, $t_{\rm A}^2$, can be obtained when one has point-to-point determinations of $T_{\rm e}$ of a given nebula. Strictly speaking, $t_{\rm A}^2$ should be considered a lower limit to $t^2$ because this last parameter, by definition, is integrated over the volume of the considered region (see further argumentation given by \citealt{2003MNRAS.340..362R} and \citealt{mesa-delgado08}).
Using this approximation and temperature measurements based on the [\ion{O}{iii}]($\lambda$4959+$\lambda5007)/\lambda4363$ line ratio, \citet{krabbe02} found
$t_{\rm A}^2\sim 0.0025$,  or equivalently  a temperature dispersion of only 5\,\% in the 30 Doradus Nebula. Other estimates of $t_{\rm A}^2$ for extragalactic SFs have derived similar low values of this parameter  \citep{tsamis03, oliveira08}. 
\citet{2003MNRAS.340..362R}, by using Hubble Space Telescope (HST/STIS) long-slit spectroscopy of the  Orion Nebula,
found $t_{\rm A}^2$ varying from $\sim$0.006 to $\sim$0.018. 
Recent radio observations in different zones of the Orion Nebula by \citet{wilson15} revealed only spatial variations in $T_{\rm e}$ of the order of 13\,\%.  Moreover, a low $t^{2}$ value of $\sim$0.005 was found from
 photoionization model simulations for SFs  by  \citet{copetti06}.

Hitherto, the $T_{\rm e}$-method  has been used to estimate the elemental abundances of oxygen and nitrogen for few  AGNs (see \citealt{alloin92, izotov08, dors15, dors17, flury20, dors20}), which are mostly based on photoionization models (see \citealt{2020MNRAS.492..468D} and references
therein). Gaseous abundance determinations in AGNs through ORLs are not available in the literature, mainly because these lines are extremely faint and are only accessible in nearby objects. In particular,  \citet{dors20} showed that O/H estimates based on $T_{\rm e}$-method in the
Narrow Line Regions (NLRs) of AGNs
are underestimated by an average factor of about 0.2 dex in comparison with those derived from
detailed photoionization models, where the discrepancy is more pronounced (reaching $\sim 0.8$ dex) for the low metallicity regime [$\rm 12+\log(O/H) \: \la \: 8.4$ or $(Z/{\rm Z_{\odot}}) \: \la \:0.5$].

In contrast to SFs, only a few studies have  addressed the electron temperature structure in AGNs. For example, \citet{revalski18a, revalski18b} found no systematic electron temperature variation along the radius of three nearby AGNs using  long slit spectroscopy. On the other hand,  electron temperature maps for the inner kpc of luminous Seyfert galaxies, derived using the [\ion{O}{iii}]($\lambda$4959+$\lambda5007)/\lambda4363$ line ratio measured from integral field spectroscopy,  show values varying from $\sim$8\,000 to  $\gtrsim$30\,000 K \citep{rogemar_te21}. These findings  demonstrate -- for the first time -- that large spatial variations of $T_{\rm e}$ are present in AGNs, providing a unique opportunity to explore the effects of $t^2$ on abundance determinations in this object class, which is a benchmark to study the properties of distant galaxies and the  high-redshift Universe.

Although temperature fluctuations have been quantified for SFs and PNs, this is the first study aimed at studying such fluctuations in AGNs. We present  a study on the electron temperature fluctuations in three Seyfert galaxies (Mrk\,79, Mrk\,348 and Mrk\,607) using 2D spectroscopy data presented by 
\citet{rogemar_te21}. The data present high-spatial and spectral resolution spectra and will provide detailed information about the spatial distribution of the physical properties of these galaxies.
In Section~\ref{datasec}, we describe the data and methodology, in
 Sect.~\ref{result} we present our results and Sect.~\ref{disc} is aimed to evaluate the electron temperature fluctuations.  Finally, the conclusions are presented in Sect.~\ref{conc}.

\section{Observational data} \label{datasec}

We use optical emission-line flux distributions to map the electron density and electron temperature in the $0.4-1.1$ kpc inner central region  of Mrk\,79, Mrk\,348 and Mrk\,607, three luminous nearby Seyfert galaxies.
These objects were selected  from \citet{Freitas18} by presenting extended [O\,{\sc iii}]$\lambda$4363 emission \citep[see][]{rogemar_te21}. 

The data were obtained with the Gemini Multi-Object Spectrograph \citep[GMOS,][]{gmos} Integral Field Unit (IFU), covering the spectral range from 4300 to 7100\,\AA\, with a velocity resolution of $\sim$90\,km\,s$^{-1}$ (estimated by measuring the full width at half-maximum, FWHM, of emission-line profiles of the CuAr calibration lamp, used to wavelength calibrate the data)  and spatial resolutions of 110--280 pc (estimated as the FWHM of the continuum flux distribution of field stars from the acquisition images). The data reduction procedure is described in \citet{Freitas18} whereby the  standard procedures using the {\sc gemini.iraf} package were followed. Here, we use the emission-line flux distributions measured by \citet{rogemar_te21}, who used the  {\sc ifscube} python package \citep{ifscube} to  investigate the variations of the electron temperature in the NLRs of the three Seyfert galaxies.  Table~\ref{tab:sample} presents the basic information of the three galaxies.

\begin{table*}
\caption{Basic information of Mrk\,79, Mrk\,348 and Mrk\,607. Columns: (1) Name of the galaxy, (2) redshift ($z$), (3) nuclear activity class, (4) bolometric luminosity ($L_{\rm bol}$) calculated using $
\log{L_{\rm bol}} = 0.0378(\log{L_{\rm X}})^2 - 2.03\log{L_{\rm X}}+61.6$ \citep{ichikawa17}, where $L_{\rm X}$ is the hard X-ray (measured in the range 14-195 keV) luminosity from \citet{BAT105}, (5) exposure time,  (6) spatial resolution estimated as the FWHM of the continuum flux distribution of field stars from the GMOS acquisition images \citep[see ][]{Freitas18}. }
\label{tab:sample}
\centering
\begin{tabular}{lcccccc}
\hline
Object    & $z$ & Nuc. Act. & log $L_{\rm bol}$  & Exp. Time & Spat. Res.    \\
          &      &           &   (erg\,s$^{-1}$)  & (sec) & (pc) \\
\hline
Mrk\,79  & 0.0222 & Sy 1  & 45.0  &  6$\times$810 & 280$\pm$30  \\
Mrk\,348 & 0.0150 &  Sy 2  & 45.3  &  6$\times$810 & 190$\pm$25  \\
Mrk\,607 & 0.0089 &  Sy 2  & 43.4  &  7$\times$810 & 110$\pm$14 \\
\hline
\end{tabular}
\end{table*}

 We recompute the electron temperatures  using the PyNeb routine \citep{luridiana15} and the $\frac{I({\rm [\ion{O}{iii}]\lambda5007})}{I({\rm[\ion{O}{iii}]\lambda4363})}$ and  $\frac{I{\rm ([\ion{S}{ii}]\lambda 6716)}}{I{\rm ([\ion{S}{ii}]\lambda 6731)}}$ emission-line intensity ratios from \citet{rogemar_te21}. For  the highest temperatures  ($T_{\rm e}\gtrsim30\,000$ K), the PyNeb based measurements are expected to be more accurate than those based on the equation from \citet{hagele08} assumed in \citet{rogemar_te21}, as the range of temperatures used to derive this equation originally does not include high temperatures.  First, we compute the electron density ($N_{\rm e}$) values for each spaxel using the [S\,{\sc ii}] line ratio and assuming $T_{\rm e}=20\,000$ K; then we calculate the electron temperature assuming the $N_{e}$ values obtained for each spaxel.  Although the [S\,{\sc ii}] emission arises from low ionization degree and partially-ionized zones and the [O\,{\sc iii}] emission traces fully ionized and perhaps higher density zones, the dependence of $T_{\rm e}$ on the adopted  $N_{e}$ appears to be negligible.

\begin{figure*}
{\centering
\includegraphics[width=0.3\textwidth]{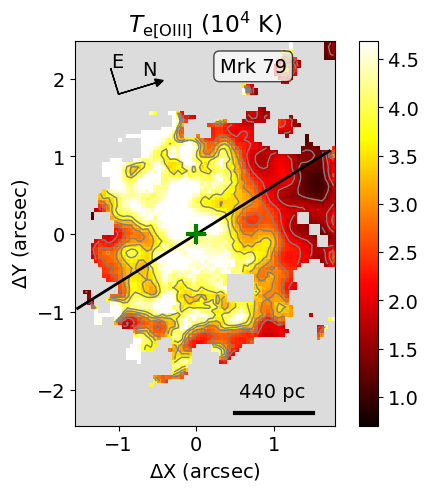}
\includegraphics[width=0.3\textwidth]{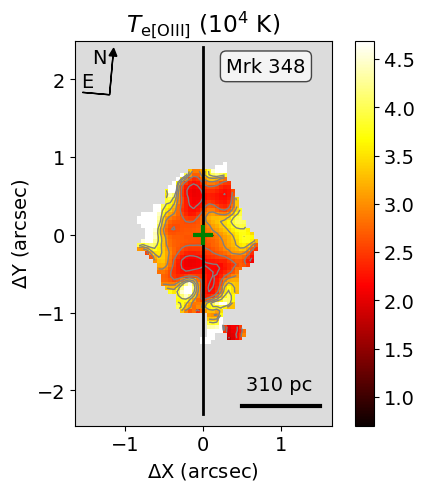}
\includegraphics[width=0.3\textwidth]{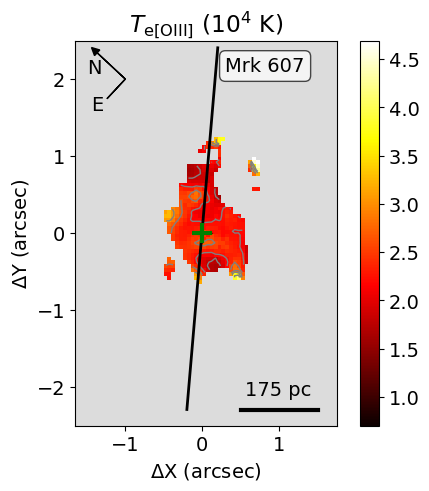}
}
\caption{Maps of the electron temperature
for Mrk 79 (left), Mrk 348 (middle) and Mrk 607 (right). The central crosses mark the position of the galaxy nuclei and the lines show the orientation of the AGN ionization axes, as obtained from [O\,{\sc iii}] HST images by \citet{schmitt03}. The physical scale and the spatial orientation of the GMOS field of view for each object are shown. The gray regions correspond to locations where the [O\,{\sc iii}]4363 emission line is not detected at a 3$\sigma$ continuum level. }
\label{maps1}
\end{figure*}

\begin{figure*}
{\centering
\includegraphics[width=0.3\textwidth]{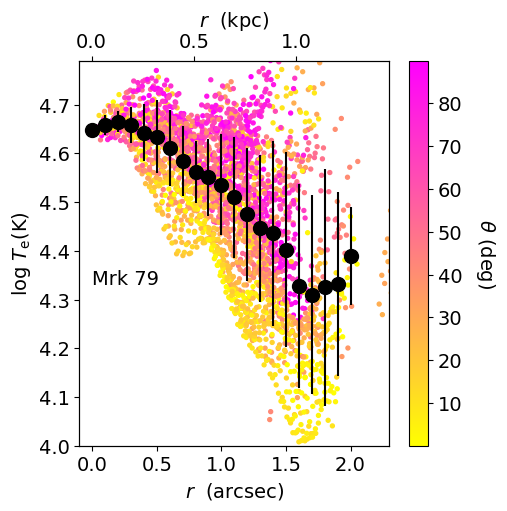}
\includegraphics[width=0.3\textwidth]{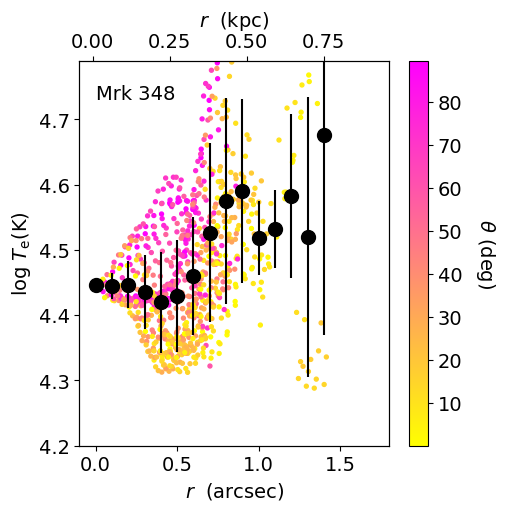}
\includegraphics[width=0.3\textwidth]{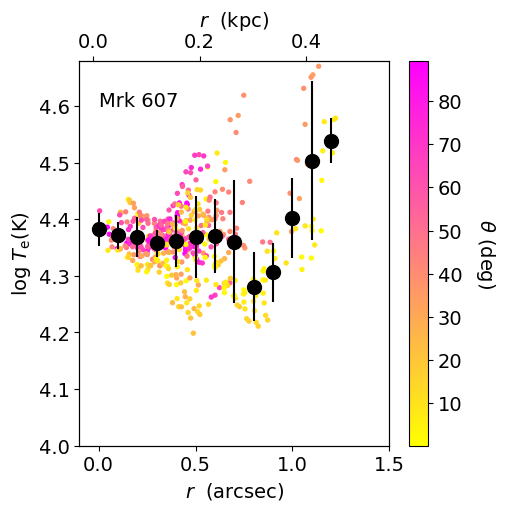}
}
\caption{Plots of the logarithm of $T_{\rm e}$ versus the distance  (in units of arcsec) from the continuum peak 
for Mrk\,79 (left), Mrk\,348 (middle) and Mrk\,607 (right). The points are color coded in according to the angle ($\theta$) between the position of each spaxel and the orientation of the AGN ionization axis (indicated in Fig.~\ref{maps1}), whose values are shown in degrees in the color bar. The filled circles represent the mean $T_{\rm e}$ values and the error bars the standard deviation of these
within bins of  0.1 arcsec. }
\label{fig:rad}
\end{figure*}

\begin{figure*}
{\centering
\includegraphics[width=0.3\textwidth]{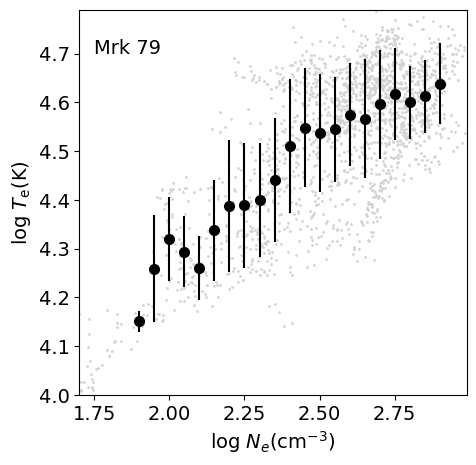}
\includegraphics[width=0.3\textwidth]{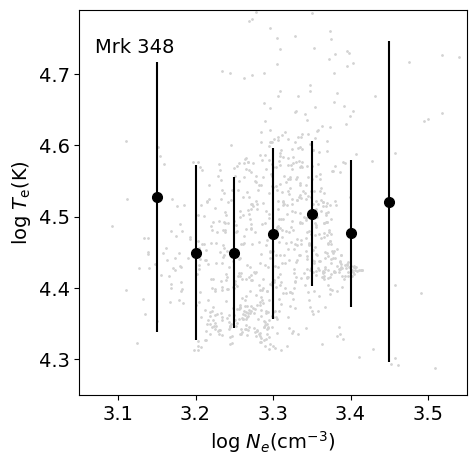}
\includegraphics[width=0.3\textwidth]{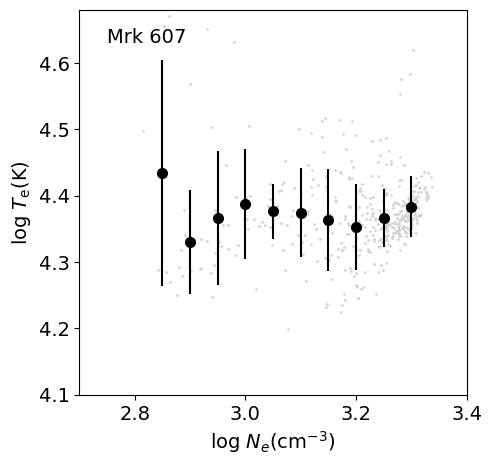}
}
\caption{Plots of  the logarithm of $T_{\rm e}$  versus  the logarithm of $N_{\rm e}$ for Mrk\,79 (left), Mrk\,348 (middle) and Mrk\,607 (right). The gray points show the values of all spaxels and the filled circles show the mean temperature values within bins of $\log N_{\rm e}/{\rm (cm^{-3})} = 0.05$. The error bars are the standard deviation of the $T_{\rm e}$ values within each bin.}
\label{fig:nete}
\end{figure*}

\section{Results}
\label{result}

The electron density maps for Mrk\,79, Mrk\,348 and Mrk\,607 were already shown in \citet{rogemar_te21}. 
 The $N_{\rm e}$ values range from $\sim$100 cm$^{-3}$ to  $\sim2\,300$ cm$^{-3}$. These values are in the range of densities reported for the central region of nearby active galaxies \citep[e.g.][]{kakkad18,Freitas18}.
The electron temperature maps for Mrk\,79, Mrk\,348 and Mrk\,607 are shown in Fig.~\ref{maps1}. These maps are consistent with those presented in \citet{rogemar_te21} based on the equation from \citet{hagele08}, but as pointed out in Sec.~\ref{datasec}, the PyNeb based values are more accurate for the highest temperature values. The $T_e$ values in these galaxies  are notably higher than those found in SFs (e.g. \citealt{kennicutt03, hagele08}) and in PNs (e.g. \citealt{2005A&A...443..981K, 2008A&A...486..545S, 2014A&A...563A..42R, 2020A&A...634A..47M}), which are in the order of 10\,000-15\,000 K. The  highest $T_{\rm e}$ values are derived in regions away from the AGN ionization axis in all galaxies. This behaviour is clearly seen in Fig.~\ref{fig:rad}, a plot of $T_{\rm e}$ versus the distance to the nucleus, where the points are color coded in according to the angle ($\theta$) between the position of each spaxel and the orientation of the AGN ionization axis (indicated in Fig.~\ref{maps1}).  The $T_{\rm e}$ values along the AGN ionization axis are similar to those obtained using long slit spectroscopy and predicted by photoionization models, i.e. in the range $\sim$10\,000 -- 25\,000 K \citep{revalski18a,revalski18b,revalski21}. 
However, the highest $T_{\rm e}$ values ($\gtrsim \: 30\,000$ K) are derived 
outside the AGN ionization axis,  originated probably by  shock ionization 
rather than photoionization \citep{rogemar_te21}.  The wide range of temperature variations derived in the galaxies studied here is consistent with the  [O\,{\sc iii}]$\lambda$4363/[O\,{\sc iii}]$\lambda$5007 map for NGC\,1068 shown by \citet{dagostino19}, which presents values in the range  $0.02 - 0.08$, corresponding to temperatures of $\sim$15\,000 -- 40\,000 K, as obtained using the PyNeb routine and assuming $N_{\rm e}=1\,000$ cm$^{-3}$.
Our results indicate  the presence of  large spatial variations of electron temperature in the gas phase of AGNs,
not observed in SFs and PNs.

 In Fig.~\ref{fig:nete} we show plots of $T_{\rm e}$  versus $N_{\rm e}$ for Mrk\,79 (left), Mrk\,348 (middle) and Mrk\,607 (right). The filled circles show the $T_{\rm e}$ values within bins of $\log N_{\rm e}/{\rm (cm^{-3})} = 0.05$. A clear correlation between $T_{\rm e}$ and $N_{\rm e}$   is observed for Mrk\,79, while no correlations are seen for Mrk\,348 and Mrk\,607. Similar correlations are found in  ring-shaped  PNs by \citet{2005A&A...443..981K}. A possible interpretation for the correlation seen in Mrk\,79 is that the lower $T_{\rm e}$ and $N_{\rm e}$, seen along the AGN ionization axis, are mainly due to photoionization by the AGN radiation, while the increase of the values of these properties with the distance to the ionization axis indicates a larger contribution of emission of compressed shock-ionized gas.

\section{Electron temperature fluctuations and implications} \label{disc}

According to \citet{peimbert69}, the temperature fluctuations over the observed gas phase volume ($V$) of a nebula can be quantified in terms of the mean electron temperature ($T_{\rm 0}$) and the $t^{2}$ parameter, which are defined as
\begin{equation}
  T_{\rm 0} = \frac{\int{T_{\rm e}N_{\rm e}N_{\rm i}{\rm d}V}}{\int{N_{\rm e}N_{\rm i}{\rm d}V}} 
\end{equation}
and 
\begin{equation}
  \label{et1}
    t^{2}=\frac{\int (T_{\rm e}-T_{\rm 0})^{2} \: N_{i} \: N_{\rm e} \: {\rm d}V}{T_{\rm 0}^{2} 
    \int  N_{i} \: N_{\rm e} \: {\rm d}V} \ ,
\end{equation}
\noindent where $N_{i}$ is the density of the ion used to measure the temperature. A direct estimate of $t^2$ along the line-of-sight  can not be obtained from our data, but we can estimate $t^2$  projected in the plane of the sky ($t_{\rm A}^{\rm 2}$) -- as well as its associated $T_{\rm 0,A}$ -- using the equations given by: 
\begin{equation}
  T_{\rm 0,A} = \frac{\sum_{j}T_{{\rm e},j}/T_{{\rm e},j}^{\rm err}}{\sum_{j}1/T_{{\rm e},j}^{\rm err} } \ 
\end{equation}
and 
\begin{equation}
  \label{et2}
  t_{\rm A}^{\rm 2} = \frac{\sum_{j}(T_{{\rm e},j}-T_{{\rm 0},A})^2/T_{{\rm e},j}^{\rm err}}{T_{{\rm 0},A}^{\rm 2}\sum_{j}1/T_{{\rm e},j}^{\rm err}} \ ,
\end{equation}
\noindent where $T_{{\rm e},j}$ and $T_{{\rm e},j}^{\rm err}$ are the electron temperature and its uncertainty in each spaxel, respectively. 
We use 1/$T_{{\rm e},j}^{\rm err}$ as a weighting factor of the point-to-point temperature instead of $N_{i}N_{\rm e}$ (as in \citealt{2003MNRAS.340..362R} or \citealt{mesa-delgado08}) for various reasons. Firstly, because $N_{\rm e}$ is determined from a [\ion{S}{ii}] line ratio, and this indicator may not be representing well the density -- and density variations -- within the O$^{++}$ zone in a huge and complex object as an AGN host. Secondly, as it has been seen in Sect.~\ref{result}, the spaxels with higher $T_{\rm e}$ values -- and with a larger contribution of shock excitation -- tend to be located in the regions away from the AGN ionization axis in all galaxies, and these correspond to the faintest zones. Since the emissivity of recombination or collisionally excited lines of typical ionized nebulae are proportional to the $N_{i}N_{\rm e}$ factor, lines affected by shock excitation are very much dependent on shock velocity and other geometrical and physical aspects of the interaction. The factor 1/$T_{{\rm e},j}^{\rm err}$ is basically proportional to the signal-to-noise ratio of the auroral [\ion{O}{iii}]$\lambda$4363 line, therefore, in this case, we think it is a better choice for weighting the contribution of each spaxel to define  $t_{\rm A}^{\rm 2}$. 

The $T_{\rm 0,A}$ and $t_{\rm A}^{\rm 2}$ values derived for the three galaxies are listed in the third and fourth columns of Table~\ref{tab:pars}.  These parameters are calculated assuming the full area with [\ion{O}{iii}]$\lambda$4363 line detected above a 3$\sigma$ level.  The electron temperature calculated by summing up the fluxes of all spaxels where the [\ion{O}{iii}]$\lambda$4363 is detected -- $T_{\rm e,sum}$ -- is also included in the second column of  Table~\ref{tab:pars}.  These quantities may be considered as representative of each entire object. 
The $t_{\rm A}^{\rm 2}$ values we find are 0.135, 0.039, and 0.015 for Mrk\,79, Mrk\,348, and Mrk\,607, respectively, implying spatial variations of $T_e$ between 12 and 37\%. The $t_{\rm A}^{\rm 2}$ found in the AGNs studied here are of the order or even much higher (up to a factor about 7) than the maximum values derived in observational studies of extragalactic SFs  \citep[e.g.][]{krabbe02, oliveira08}, the Orion Nebula \citep[e.g.][] {2003MNRAS.340..362R, mesa-delgado08} or PNe  \citep[e.g.][]{2005A&A...443..981K, 2014A&A...563A..42R}. Our results suggest that the potential impact of $t^2$ on abundance determinations based on the $T_{\rm e}$-method may be important for some AGN hosts.

\begin{table}
\caption{Electron temperature ($T_{\rm e,sum}$) calculated by summing up the fluxes from the spaxels with $T_e$ measurements, mean electron temperature ($T_{{\rm 0},A}$), and surface temperature fluctuation ($t_{\rm A}^{\rm 2}$) for the three objects. The uncertainties in the temperatures results from propagating the 1$\sigma$ error of the  [\ion{O}{iii}]$\lambda5007$/$\lambda4363$ flux line ratio.   }
\label{tab:pars}
\centering
{\renewcommand{\arraystretch}{1.2}
\begin{tabular}{lccc}
\hline
Object    &   $T_{\rm e,sum}$  &  $T_{\rm 0,A}$ &  $t_{\rm A}^{\rm 2}$ \\
     & ($10^{4}$ K) &  ($10^{4}$ K) &    \\
\hline
Mrk\,79  & 3.74$_{-0.35}^{+0.49}$ &  3.00$_{-0.37}^{+0.59}$    & 0.135 \\
Mrk\,348 & 2.88$_{-0.29}^{+0.43}$ &  2.79$_{-0.37}^{+0.38}$   & 0.039 \\
Mrk\,607 & 2.40$_{-0.17}^{+0.21}$ &  2.36$_{-0.24}^{+0.34}$   & 0.015 \\
\hline
\end{tabular}
}
\end{table} 

\section{Conclusions }
\label{conc}

We have studied the temperature fluctuations in the  inner  0.4--1.1 kpc region of three nearby Seyfert galaxies: Mrk\,79, Mrk\,348, and Mrk\,607. In particular, for Mrk\,79, an AGN with 
strong  ionized outflows, we derive a direct relation between electron temperature and electron density, 
possibly caused by compressed shock-ionized gas. 
The temperature fluctuations are quantified by the
$t_{\rm A}^{\rm 2}$ parameter,  which is derived in the range of 0.015 --  0.135, of the order or even larger than the maximum values obtained for galactic and extragalactic star forming regions (\ion{H}{ii} regions) and for planetary nebulae. Therefore, the effect of temperature fluctuations may be important in gas phase abundance estimates 
in AGNs based on $T_{\rm e}$-method.

\section*{Acknowledgements}
We thank an anonymous referee for their critical reading and suggestions that helped us to improve our paper.
R.A.R. acknowledges support from Conselho Nacional de Desenvolvimento Cient\'ifico e Tecnol\'ogico (CNPq) and Funda\c c\~ao de Amparo \`a Pesquisa do Estado do Rio Grande do Sul (FAPERGS). O.L.D. and A.K. acknowledge support from CNPq  and    Funda\c c\~ao de Amparo \`a Pesquisa do Estado de S\~ao Paulo (FAPESP).
 C.E. acknowledges support from the Agencia Estatal de Investigaci\'on del Ministerio de Ciencia e Innovaci\'on (AEI-MCINN) under grant {\it Espectroscop\'\i a de campo integral de regiones \ion{H}{II} locales. Modelos para el estudio de regiones \ion{H}{II} extragal\'acticas} with reference 10.13039/501100011033.
Based on observations obtained at the Gemini Observatory, which is operated by the Association of Universities for Research in Astronomy, Inc., under a cooperative agreement with the NSF on behalf of the Gemini partnership: the National Science Foundation (United States), National Research Council (Canada), CONICYT (Chile), Ministerio de Ciencia, Tecnolog\'{i}a e Innovaci\'{o}n Productiva (Argentina), Minist\'{e}rio da Ci\^{e}ncia, Tecnologia e Inova\c{c}\~{a}o (Brazil), and Korea Astronomy and Space Science Institute (Republic of Korea).
\section*{Data Availability}
The data used in this paper is available in the Gemini Science Archive under the project code GN-2014B-Q-87.

\bibliographystyle{mnras}
\bibliography{paper_final} 







\bsp	
\label{lastpage}
\end{document}